\documentclass[%
 reprint,
 aps,
prab,
]{revtex4-2}

\usepackage{bm}                       
\usepackage{color}    			      
\usepackage{dcolumn}                  
\usepackage{graphicx}                 
\usepackage{hyperref}			      
    \hypersetup{colorlinks = true,    
                 linkcolor = blue,    
               linktocpage = true,    
               anchorcolor = blue,    
                 citecolor = blue,    
                 filecolor = red,     
                 menucolor = red,     
                  runcolor = red,     
                  urlcolor = blue,    
				}

\usepackage[T1]{fontenc}
\usepackage{setspace}     
\usepackage[separate-uncertainty=true,multi-part-units=single]{siunitx}

\begin{document}
\newcommand{\cste}{Cs\textsubscript{2}Te}

\title{European Workshop on Photocathodes for Particle Accelerator Applications 2022:\\Summary Report}

\author{L. Monaco}
\author{D. Sertore}
    \affiliation{INFN Milano -- LASA, Via Fratelli Cervi 201, Segrate (MI), Italy}

\author{M. Baylac}
    \affiliation{Univ. Grenoble Alpes, CNRS, Grenoble INP, LPSC-IN2P3, 38000 Grenoble, France}

\author{L.B Jones}
\author{T.C.Q. Noakes}
    \affiliation{ASTeC, STFC Daresbury Laboratory, Warrington, Cheshire, WA4 4AD, UK}
    \affiliation{Cockcroft Institute of Accelerator Science \& Technology, WA4 4AD, UK}

\author{J. K\"uhn}
    \affiliation{HZB, Helmholtz--Zentrum Berlin für Materialien und Energie GmbH, 14109 Berlin, Germany}

\author{R. Xiang}
    \affiliation{HZDR, SRF--Gun Group, ELBE Department, Institute of Radiation Physics, Helmholtz--Zentrum Dresden--Rossendorf, 01328 Dresden, Germany}


\begin{abstract}
\end{abstract}

\maketitle

\onecolumngrid

\begin{figure}[]
    \label{GroupPhoto}
        \vspace{16pt}
            \includegraphics[width=\textwidth]{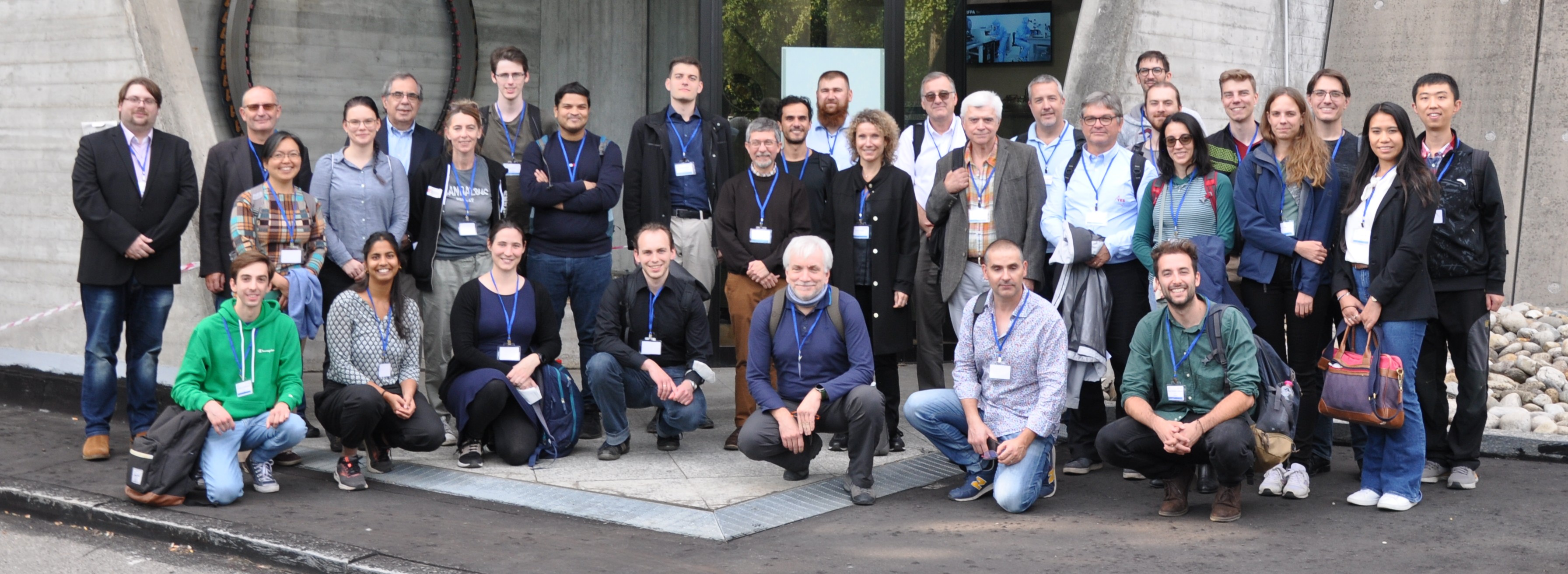}
                \begin{center}
            Delegates at the EWPAA 2022 Workshop during their visit to the LASA facility
        \vspace{32pt}
    \end{center}
\end{figure}

\twocolumngrid


The European Workshop on Photocathodes for (particle) Accelerator Applications (EWPAA) brings together experts in the field of photocathode based electron sources for use in particle accelerators, with the aim of sharing their knowledge and latest research and development progress in this crucial field of particle accelerator science.  

The workshop is convened every other year, and is thus complementary to the P$^3$ workshop (Photocathode Physics for Particle accelerators) run in the USA.  Consequently, there is a workshop focusing on photocathodes for particle accelerator applications convened every year, either in Europe or the USA.

The EWPAA 2022 is the 4\textsuperscript{th} meeting in this workshop series.  The event was hosted by the INFN LASA Institute in Milan in collaboration with University of Milano between September $20^{\mathrm{th}}$ and $22^{\mathrm{nd}}$.  Details of the event and the scientific programme can be found at the website:

\begin{center}
    \href{https://agenda.infn.it/event/ewpaa2022}{https://agenda.infn.it/event/ewpaa2022}
\end{center}

The programme was organised with 7 working groups, with each oral contribution assigned to the most appropriate group.  This report presents, summarised by the work group conveners, the innovative ideas, the challenges and the main points raised by each of the speakers.

    \linespread{1.5}
        \tableofcontents
    \singlespacing

\section{\label{sec:WG1}WG1:Overview of Photocathode Research}

Working Group 1 was focused on general overviews of key areas of photocathode R\&D.  It was convened by Daniele Sertore.

\subsection{\label{WG1-1}High Brightness Beams}

Alberto Bacci (INFN LASA, Italy) presented a view to photocathodes seen from beam dynamics applications with particular attention to the achievement of high brightness. Within the photocathode desiderata, high Quantum Efficiency (QE) and fast response time were the most important to achieve final requested beam performances. Linked to the photocathode requirements, laser beam transverse shaping plays a key role. It was shown that a truncated Gaussian delivers a smaller final emittance w.r.t. a uniform distribution, as usually thought. The concepts before presented were applied to three different scenarios, namely BRiXSinO ERL project based on a DC gun, the MariX project based on a RF gun "Apex--like" and finally the EuPRAXIA configuration with a 1.6 cells S--band RF Gun. BriXSinO is based on a novel scheme where the beam pass is accelerated through a Superconducting Linac, is bent in a compressing arc and then re--injected into the linac from the exit side. The SC linac can be then used to decelerate the beam in an ERL configuration, or to accelerate the beam in the usual way in the so call Two--Pass Two--Way operation mode. The MariX project was developed as a possible accelerator facility to be installed in the new Milan university campus called MIND. A full start--to--end simulation of the accelerator is presented with emphasis on the optimization of the injector with a self--developed genetic algorithm GIOTTO to achieve optimal compression in the MariX Loop and minimal transverse emittance for FEL operation. For the last example, EuPRAXIA, Alberto presented preliminary results on the optimization of the injector in order to achieve the best performance from the beam driven plasma acceleration experiment.

\subsection{\label{WG1-2}Recent Advancements in Photocathodes}

Laura Monaco (INFN LASA, Italy) gave an overview of the recent photocathode advancements. In the field of metallic photocathodes, good results have been obtained at HZDR in the SRF gun with Magnesium. Studies are also on--going to lower the metal work function by depositing MgO film or exploring single vs polycrystalline Ag or even creating structure on top of a flat metallic film to enhance photoemission. Turning to \cste{} semiconductor photocathodes, good performance levels were reported with operating lifetime more than 1,400 days while conserving good QE, and with response time in the \SI{200}{\femto\second} range. 
The cesium telluride photocathode has been also used extensively in the HZRD SRF Gun, even if degradation of the film is still a not solved problem. 
Co--deposition of \cste{} film has been studied by a joint experiment CERN--STFC as well as at STFC directly for the CLARA project. Alkali antimonide photocathodes are, nowadays, the more extensive studied photocathodes due to their sensitivity to visible light and high QE. INFN--DESY PITZ collaboration has tested a set of KCsSb photocathodes, deposited at INFN LASA, in the PITZ RF gun. This experiment showed a larger field emission of antimonide w.r.t. telluride but also a very fast response time (below \SI{100}{\fs}) and a small thermal emittance. At PITZ, there was a confirmation of the sensitivity of these kind of photocathodes to vacuum conditions as well as exposure to high accelerating gradients (above \SI[per-mode=symbol]{40}{\mega\volt\per\metre}). 
A similar experiment has been done in a joint collaboration between BNL and UCLA using a NaKCs photocathode. In this case, the QE degraded consistently during the transportation. Nevertheless, the team was able to measure QE, MTE and non-linear photoemission effects.
Also at HZB, NaKSb is studied and characterized in preparation to its usage in the HZB \SI{1.3}{\GHz} SRF Gun. 
There is significant on--going R\&D activity in China on a broad set of photocathodes and growth techniques. 
Moving to Cs\textsubscript{3}Sb material, the growth of an epitaxial film of Cs\textsubscript{3}Sb(100) on a 3C Si(100) has allowed, for the first time, ARPES measurement of the band structure of this material. The choice of a lattice match material like strontium titanate has improved the smoothness of the film and hence the contribution to MTE from roughness. 
On the way to improved alkali antimonide robustness, Laura presented the results obtained at LANL on graphene as substrate or graphene as protective layer for the Sb film. Both of them show interesting results that need further studies. 
To improve QE, the idea was presented of using Surface Plasmon Polaritons to enhance the laser absorption by minimizing the reflection at a specific wavelength.
The last part of the presentation focused on new photocathode test facilities, important tools for developing and testing new ideas and processes. Finally, the Machine Learning and Artificial Intelligence approach to photocathode material selection and photocathode growth closed the presentation.

\subsection{\label{WG1-3}Experience on Polarized Photocathodes}

Kurt Aulenbacher (University of Mainz, Germany) presented on `polarized' photocathodes. After a short introduction of Negative Electron Affinity Photocathode (NEA) characteristics, Kurt introduced the well know GaAs photocathodes, largely used for generation of spin polarized electrons, in the strained and superlattice configuration, necessary for enhancing the electron beam polarization. A new entrance in this family is the "resonant" strained superlattice cathode that make use of an enhanced absorption achieved by a DBR--Reflector that causes the active region to act as a cavity with enhanced absorption at resonance. This type of photocathodes showed high QE and high polarization in the \SI{770}{\nm} region. Moreover they should be fast, not affected by `photovoltage' effect and able to deliver \si[per-mode=symbol]{\ampere\per\square\cm} while having small thermal emittance.
Polarized beams are important for nuclear an particle physics experiment. Two machines are nowaday running with polarized electrons, namely MAMI in Germany and CEBAF in USA. MESA (ERL), that is in construction, will also operate with these beams. Future machines that aim to use polarized beams are, among others, LHeC, CERL but they might be challenged by technical and physical limitations of these material, because they require extraction of \SIrange{10}{100}{\kilo\coulomb}.
Changes in surface layer will strongly affect the photocathode performance. Chemical reactions with gases of the environment, ion back bombardment and thermal decomposition are the most important players in this context. To mitigate them, UHV or XHV vacuum, very limited beam losses in the source region and cathode temperature below \SI{50}{\degreeCelsius} are required.
A different approach is to stabilize the surface or protect it. Different labs have explore different solutions: TU Darmstand \& JLAb use Cs:F:Li while Cornell has explored Cs:O:Sb and \cste{}. These are still in the R\&D phase bu the results are encouraging.
To control the cathode temperature, in the "BETHE" project a new approach is presented where the puck is pressed against a thermal conducting insulator like Boron Nitride (BN). This operation is done before moving the cathode into the insulator. Given the properties of BN, this solution should guarantee reasonable operation of the cathode also with large incident laser power.
Besides the technical limitation, a physical limitation is the surface photovoltage, i.e. modified NEA--state reduce the QE due to near band-gap excitation. This has significant impact of photocathode performances.
The last point touched in this presentation is the lack of a commercial supplier for the lattice strained photocathode. This requires a common effort between laboratories and institutions to support the manufacturing of these fundamental photocathodes needed for all modern applications requiring polarized electron beams.

\section{\label{sec:WG2}WG2: Photocathode Performance in Accelerator Applications}

Working Group 2 was convened by Rong Xiang and focused on photocathode performance in accelerator applications.  It remains a goal in photocathode R\&D to improve the cathode performance in accelerator applications. On one hand, we need to further push the study of mature photocathodes; on the other hand, we are searching new materials with high QE, long time, low dark current, fast response time, especially low Mean Transverse Energy (MTE) for special users (hard X--ray FEL and microscope).  

\subsection{\label{WG2-1}JLAB DC Gun Developments}

Carlos Hernandez--Garcia (Thomas Jefferson National Accelerator Facility, USA) discussed the successful application of photocathodes in different type of electron sources (DC guns, NC RF guns and SRF guns).  JLab's experience of DC guns with GaAs photocathodes represents a good example. With the now mature GaAs photocathode, the CEBAF DC gun can provide highly polarized electron beam at \SI{130}{\keV} and  \SI{0.2}{\mA} CW for nuclear physics experiments for over a decade. Recently JLab team has demonstrated up to \SI{300}{\keV} and \SI{5}{mA} CW un--polarized beam for ERL. 

\subsection{\label{WG2-2}CsK\textsubscript{2}Sb Photocathodes for Application in an Industrial Accelerator}
 
Recent years have seen several new/emerging electron gun projects. In addition, the stability and reliability of photocathodes in laboratories has pushed their application in industrial accelerator projects.  Veronica Kuemper (Research Instruments, Germany) spoke about work by Research Instruments as part of the Lighthouse project, highlighting the promising application of two \SI{350}{\kV} DC guns with a CsK\textsubscript{2}Sb preparation system has the goal to provide \SI{40}{\mA} in CW mode for continuous operation 23h/7 days a week.

\subsection{\label{WG2-3}Status of Precise Measurements of Electron Beam Polarization Changes During Long Term Operation}

Currently the mature semiconductor cathodes ready for various applications include \cste, Cs\textsubscript{2}KSb and GaAs. Jennifer Groth (University of Mainz) shared data from MESA on their use of GaAs photocathodes in DC guns realize high polarization during long--term operation, highlighting their observations thtat preparation with NF\textsubscript{3} allows more activation cycles with stable QE. 

\subsection{\label{WG2-4}Photocathode Activities at DESY}

Groups operating photoinjectors have accumulated rich experience during their beam operation. Firstly, extremely high vacuum (XHV) environment is the key for long operation lifetime in every electron gun. Furthermore, it is clear that the ion back--bombardment is responsible for the cathode surface damaging during DC gun operation. However, further studies are needed to confirm that ion back--bombardment is detrimental in RF guns.
David Pavel Juarez--Lopez (DESY, Germany) presented recent results on the successful application is the \cste{} cathode for \SI{1.3}{\giga\Hz} normal conducting RF guns at FLASH and European XFEL, which work reliably at DESY for years already and fit well for FLASH and European XFEL as a user facility.

\subsection{\label{WG2-5}Performance of Bialkali Photocathode in DC--SRF Photoinjector}

Haumu Xie (Peking University, China) shared their experience of  DC--SRF gun operation, noting that the cryogenic temperature has strong influence on the QE of Cs\textsubscript{2}KSb, which is lower at cryogenic temperature than that at room temperature.  The bialkali photocathode in the DC--SRF--II gun has already produced the first beam, with typical QE around \SI{1}{\percent} at \SI{36}{\kelvin}. 

For the first goal, the analysis of cathode after gun operation without explosion to air will help us to understand the cathode surface changing or QE distribution evolution during gun operation.  And for the second goal, a real test platform (test gun) is important to prove the cathode properties suitable for the gun operation. 

\subsection{\label{WG2-6}Photocathode Performance at LCLS--II}

INFN style \cste{} photocathodes have been used in many large facilities for decade. In last years, \cste{} cathodes were successful prepared and operated in VHF gun for LCLS-II project too. Theo Vecchione (SLAC, USA) gave an update on progress in commissioning the LCLS--II VHF gun, and their use of \cste{} cathodes.  He drew attention to their work to reduce dark current, highlighted the importance of post--mortem cathode evaluation, and the particularly the preservation of the used cathode state through the use of in--vacuum transfer to the post--mortem analysis facility.  He also gave an overview of SLAC's work to construct a cryogenically--cooled momentatron system to measure MTE.

\section{\label{sec:WG3}WG3: New Photocathode Ideas}

Working Group 3 was focused on new ideas and concepts within the field of photocathode technology.  It was convened by Laura Monaco.

\subsection{\label{WG3-1}Ultra--thin MgO Films on\\Metal Photocathodes to Enhance QE}

Chris Benjamin (STFC Daresbury Laboratory, UK) presented work on his application of ultra--thin oxide films to reduce work function and so increase the photocathode QE.  Metal photocathodes are used for their ease of use and relative robustness to less than ideal vacuum conditions. However, their high work function and consequent low quantum efficiency is a hindrance in their utilisation as high beam currents require high power levels of UV light. The application of an ultra--thin MgO film enhances the quantum efficiency by nearly an order of magnitude on both Ag(100) and Cu(100) single crystals by reducing the work function of the metal cathode.  At the same time, the robustness of the photocathode to residual vacuum gases is improved.  The chemical nature of this modified surface was investigated through X--ray photoelectron spectroscopy (XPS), low energy electron diffraction (LEED) and quantum efficiency measurements which confirmed an increase in electron yield.  Measurements of the mean transverse energy (MTE) using the TESS system at Daresbury show that the photoemission threshold is extended through the deposition of an ultra--thin oxide layer, and also that the energy spread of the photoemitted electrons agrees well with the Dowell--Schmerge approximation. 

\subsection{\label{WG3-2}Plasma Photocathode}

Carlo Benedetti (LBNL, USA) presented an update on progress in the field of plasma photocathodes.  Plasma--based \textit{accelerators} have received significant interest in recent years owing to their ability to sustain large acceleration gradients several orders of magnitude larger than in conventional RF accelerators (\SIrange[range-units=single,per-mode=symbol]{10}{100}{\giga\volt\per\metre} demonstrated experimentally), enabling compact accelerating structures for future high--energy applications (e.g., particle colliders) and radiation production (e.g., X--ray FELs). In a plasma accelerator a drive beam with suitable duration and intensity (either a laser pulse in a laser wakefield accelerator, or a particle bunch in a plasma wakefield accelerator) propagating in a plasma excites a plasma wave (or wakefield). The plasma wave has a relativistic phase velocity, a characteristic size of $\sim$\SIrange[range-units=single]{10}{100}{\micro\meter} (i.e., on the order of the plasma wavelength, which depends on the background plasma density), and can accelerate and focus a properly delayed particle beam. For instance, laser wakefield accelerators have demonstrated the production of high charge (\SIrange[range-units=single,range-phrase=--]{10}{100}{\pico\coulomb}), short (\SIrange[range-units=single,range-phrase=--]{1}{10}{\femto\second}), and high--quality (<\,1\,$\mu$m normalized emittance with $\sim1$\,\% relative energy spread) electron bunches, resulting in a 4D brightness of $\sim10^{18}$\,A/mrad\textsuperscript{2}. These results were obtained in regimes where bunch production and acceleration was done by means of a single laser driver. 

In the \textit{plasma photocathode} scheme the production and acceleration of the electron bunch are decoupled. The (laser or particle) driver is responsible for the creation of the plasma wave while bunch generation is achieved by means of a short--wavelength, low--intensity, trailing ionization pulse which liberates tunnel--ionized electrons from a high--Z gas (not fully ionized by the driver) directly into the plasma wave. Modeling shows that the plasma photocathode concept allows, in principle, the production of electron bunches with a normalized emittance of $\sim$\SI{10}{\nm} which results in a significant increase of the 4D brightness up to $\sim10^{20}$\,A/mrad\textsuperscript{2}. Experiments to demonstrate the plasma photocathode concept are underway at FACET (SLAC) and the BELLA Center (LBNL).

\subsection{\label{WG3-3}CERN Photocathode Activities}

Miguel Martinez--Calderon (CERN) gave an update on photocathode activities at CERN, focusing on the fabrication and performance of\cste{} cathodes for use in the CERN Electron Accelerator for Research (CLEAR) and the Advanced proton--driven plasma Wakefield acceleration Experiment (AWAKE). For the CLEAR facility, CERN has implemented an in--situ solution utilising two evaporators within the RF photoelectron gun in an isolated small preparation chamber. This allows the cathode to be rejuvenated without opening the gun or exchanging the cathode plug through the sequential evaporation of a Tellurium layer and a Cesium layer, resulting in initial quantum efficiency (QE) values of up to \SIrange[range-units=single,range-phrase=--]{2}{3}{\percent} and sustained QE values over \SI{0.5}{\percent} during year--long runs. \cste{} photocathodes for the AWAKE experiment are fabricated and tested at CERN's photoemission laboratory using the co--evaporation technique, achieving measured post--fabrication QE levels up to \SI{24}{\percent}  in the same laboratory in a test \SI{65}{\kilo\volt} DC electron gun. These photocathodes are then transferred to the AWAKE RF photoelectron gun using a UHV transport carrier, and during AWAKE 2021 -- 2022 runs, the QE values have been shown to reach a maximum of \SI{25}{\percent} and a consistent performance.

In addition, Miguel presented CERN's latest research studies on ultrafast laser material surface nanopatterning as an alternative to improve the photoemissive properties of metallic photocathodes. By tailoring the physical dimensions of these surface nanostructures, one can localize the optical field intensity and exploit plasmonic effects occurring in such nanostructures. As a result, this surface nanopatterning technique can become a great tool for improving metallic photocathodes photoemission behavior enabling their use for next generation high brightness electron sources. Miguel presented the performance of two different femtosecond laser nanopatterned plasmonic copper photocathodes, analyzed by measuring the quantum yield with a \SI{65}{\kilo\volt} DC electron gun utilizing \SI{266}{\nm} laser excitation generated by a nanosecond laser with \SI{5}{\ns} pulse duration and \SI{10}{\hertz} repetition rate. By comparing the electron emission of the copper surface nanostructured areas with that of a flat area, their results suggest quantum yield enhancements of up to 5 times.

\subsection{\label{WG3-4}Protective Layers on Bi--alkali Cathodes}

Nathan Moody (LANL, USA) presented work on the use of 2D protective coatings for photocathodes.  The aim of this project is to address a recognised accelerator R\&D priority which is the development of a photocathode technology which simultaneously provides long operational lifetime, high efficiency and high brightness.  The solution under investigation involves the use of a 2D material which acts as a membrane to protect the photoemissive surface from exposure to pollutants.  Work published in 2017 demonstrated the concept for metals.  It showed that that a graphene layer on a copper photocathode conferred an improvement of 8 orders of magnitude in operating pressure whilst maintaining QE.  The next phase of this project focused on using a 2D material to protect a bi--alkali photocathode.  XRD was used to to demonstrate a high level of K\textsubscript{2}CsSb crystallinity following deposition onto graphene, nickel and stainless steel substrates, and XRF to confirm near--ideal stoichiometry (K\textsubscript{1.85}Cs\textsubscript{1.08}Sb).  High resolution QE maps showed that graphene was the best protective layer with the QE of the deposited film being uniform and typically \SIrange[range-phrase=--,range-units=single]{15}{17}{\percent}.  The latest stage of this project has shifted attention to studying the effect of the thickness of the 2D graphene layer on the optical performance and QE of the photocathode structure.  Measured transmission through 2, 3 and 5 layers of graphene were consistently lower than that predicted by theory:  transmission of \SI{5}{\percent} measured at \SI{5}{\eV} through 2 layers of graphene compared to \SI{50}{\percent} predicted by theory.  It has emerged that graphene is the most resilient protective layer in that the difference between the QE for a `new' and a `used' graphene surface is very small, while that for silicon and molybdenum is much larger.  QE is therefore more reproducible and resilient on graphene when reused.  It was also noted that the deposition of a bi--alkali layer on a graphene--coated metal substrate serves to increase the QE by enhancing the `mirroring' effect whereby incident photons not initially absorbed are reflected from the metal surface can still drive photoemission on their second pass through the graphene/bi--alkali photoemissive layer.  Future work will focus on the use of hexagonal boron nitride instead of graphene as the protective layer as theory predicts higher levels of QE with the same level of protection.

\subsection{\label{WG3-5}AI/ML--selection of Air--stable Photocathodes}

Evan Antoniuk (Stanford University \& LLNL, USA) gave an overview of progress in the application of machine learning to screen potential photocathode source candidate materials.  Despite considerable research effort, the discovery of new photocathode materials has historically proven to be challenging. Photocathodes must meet a vast array of material design requirements and the experimental characterization of photocathode materials is time intensive. Recent advances in computational modelling, screening methodology, and machine learning offer a unique opportunity to overcome these challenges by allowing for hundreds of thousands of materials to be simultaneously screening against multiple selection criteria.  
 
The team have carried out the first such data--driven screening to identify novel photocathode materials from a list of more than 74,000 candidates. Notably, the search space of this effort is over 3 orders of magnitude larger than what has been experimentally explored to date and encompasses a diverse spectrum of candidate materials. In addition, we utilize machine learning predictions of work functions of  90 ,698 surfaces to solve the problem of identifying low work function ($< 3$\,eV) photocathode materials.  Two distinct screenings are performed to identify photocathodes which meet two sets of selection criteria. First they identify novel high brightness photocathodes, resulting in the discovery of candidates that exhibit intrinsic emittances that are up to 4x lower than currently used photocathodes. Then they identify air--stable photocathode materials to overcome the extreme air--sensitivity of current photocathodes. This screening identifies the air--resistant photocathodes (M\textsubscript{2}O, M=Na,K,Rb) that are also visible--light photosensitive. Finally, they extensively validate the performance of these materials by utilizing previous reports in the literature and state--of--the--art simulation capabilities. 
 
The high brightness and air--resistant photocathode materials discovered in this work should pave the way for the experimental realization of higher brightness and longer lifetime photonic devices.  Their vast photoemission dataset created in this work also provides statistical insights into engineering better photocathode materials.

\section{\label{sec:WG4}WG4: Metallic Photocathodes}

Working Group 4 looked at the performance of metal photocathodes in accelerator environments, and also at work to improve their performance.  It was convened by Lee Jones.

\subsection{\label{WG4-1}Studies on the Evolution of MTE for Photocathodes Subjected to Controlled Degradation by Gas Exposure}

Liam Soomary (University of Liverpool, UK) presented work on the degradation of polycrystalline copper photocathodes, and the performance of caesium--implanted copper photocathodes.  The degradation work involved use of the Transverse Energy Spread Spectrometer (TESS) to monitor the evolution of the photoemission footprint from a copper photocathode under illumination at \SI{256}{\nano\metre} which was subjected to continuous exposure to a number of typical gas species found in the residual atmosphere in a vacuum system.  The work showed that following an initial decline in QE of 6 percentage points over the first 10\,Langmuirs of oxygen exposure, the QE remained largely stable at a level of \SI{94}{\percent} of the initial QE over an exposure of 100\,Langmuirs.
Work has also been done to manufacture and characterise polycrystalline copper photocathodes implanted with caesium ions.  The implantation was carried out in two stages using a Hiden IG5C Cs surface ion source.  The first stage involved implantation at an energy of \SI{500}{\eV}, and the second at \SI{300}{\eV}.  When characterised using the TESS over a range of wavelengths, it was found that the caesium implanted copper photoemitted over broad range of wavelengths from \SI{236}{\nano\metre} to more than \SI{500}{\nano\metre}.  Energy spread measurements with the TESS show that the \SI{25}{\milli\eV} thermal floor was reached for the \SI{500}{\eV} implanted cathode at an emission threshold around \SI{495}{\nano\metre}, increasing to around \SI{525}{\nano\metre} after the second implantation at \SI{300}{\eV}.  Both data sets fit well with the Dowell--Schmerge model with workfunctions of 2.2 and 2.3\,eV respectively.
The photoemission images form TESS indicated the presence of two different photoemission sources when illuminated at short wavelengths, the dominant low--MTE source being attributed to copper and the weaker high-MTE source driven by the presence of caesium.  Degradation studies showed that this composite photocathode is more robust on exposure to oxygen than GaAs by a factor of around 10.

\subsection{\label{WG4-2}Copper Photocathodes for CLARA}

In this presentation, Tim Noakes (STFC Daresbury Laboratory, UK) reviewed the current status of the CLARA accelerator.  He also updated on current phase 2 development work, the Full Energy Beam Exploitation (FEBE) and aspirations for phase 3.  He also reviewed the performance of the Alpha\,X \SI{10}{\hertz} RF gun and the history of the solid copper backplate photocathodes used therein, before continuing to discuss progress with the newly--designed 2.6 cell S--band gun.
The new gun uses a molybdenum INFN--style photocathode puck as its baseline.  The gun was conditioned using such a cathode, but a high level of dark current was seen which was attributed to the high levels of measured surface roughness of this cathode (c. \SI{113}{\nano\metre}).  The gun was also operated using thin film copper was deposited onto a molybdenum cathode puck by magnetron sputtering.  Though initial QE measurements were favourably high, the lifetime was very poor.  
A hybrid photocathode assembly with a bulk copper tip mounted in a molybdenum body was also investigated.  Initial tests were not promising, but following a series of in--accelerator tests a hybrid photocathode and preparation process were reached.  This involved the use of BPS172 to clean the copper, on--axis and off-axis diamond turning with a notable reduction in surface roughness of the copper tip, dimensional adjustment to move the position of the photocathode surface in the RF cavity (affecting the surface field), and finally argon plasma treatment.  The resulting hybrid photocathode demonstrated an operational 1/e lifetime of 270 days, with an acceptable level of dark current around \SI{250}{\pico\coulomb}.  This cathode operated for 6 months during the last exploitation period.
The next steps will be the commissioning of a new high repetition rate 1.5 cell \SI{400}{\Hz} S--band gun.  The photocathode choice for this gun will benefit from the previous experience with the hybrid cathode for the 10 Hz gun, and also from the use of CsTe photocathode currently under development at Daresbury.

\subsection{\label{WG4-3}Copper photocathode experience at FERMI}

Mauro Trov\`{o} (Elettra Sincrotrone Trieste, Italy) opened this talk by highlighting the demands placed on the FERMI gun through 6,000 hours per year of FEL operations.  The FERMI @ Elettra facility uses a 1.6 cell RF gun.  It incorporates a cooled backplate but does not have a cathode exchange mechanism.  Radiabeam Technologies manufactured the gun, which has been operational since May 2013.
Gun chamber pressure logging has demonstrated that the pressure rose progressively through 2020 from \SI{4E-10}{\milli\bar} to \SI{4E-9}{\milli\bar}.  QE data for 2021 shows several steps in QE, with a general downward trend which is most likely linked to the gradually increasing pressure.  The cathode was replaced in August 2021, and although this new cathode exhibited a lower QE than the previous one, the QE level has remained fairly consistent during operations through to July 2022.  However, differences have been noted between the virtual cathode laser image and the extracted beam image which had very different Twiss parameters and necessitated an adjustment of the machine optics.  Differences in photocathode QE may in general impact machine optics and prompt retuning.
Dark current levels up to 1 nC during the RF pulse without the presence of a drive laser beam were noted, and while only a small proportion of these electrons are transported any distance, they present a problem to the machine protection system.  ICT measurements suggest that the integrated dark current over the course of 1 RF pulse is approximately equal in terms of total charge to the actual electron beam pulse, emphasising the need to avoid dark current generation in the first instance, and the requirement for a robust and stable photocathode electron source.

\subsection{\label{WG4-4}Technology Aspects in Preparation at NCBJ of Nb--Pb Superconducting Photocathodes for XFEL--Type RF Electron Guns}

Jerzy Lorkiewicz (Narodowe Central Bada\'{n} J\k{a}drowych, Poland) first summarised the design parameters for the \SI{1.3}{\giga\Hz} RF gun for an XFEL and the PolFEL.  A Pb/Nb thin film has been selected to achieve this due to compatibility with the superconducting gun, durability to surface contamination and laser-induced damage, its long lifetime, and its short Cooper pair recombination time which facilitates repetition rates up to \SI{100}{\kilo\Hz}.
The photocathode preparation process was outlined, and the choice of cathodic arc Pb deposition justified through data showing that the required QE was achieved in this way.  The components of intrinsic emittance were identified as those arising from thermal emittance, surface roughness and surface field.  This emphasised the need to minimise surface roughness. 
Two cathodic arc deposition methods were presented:  Filtered Arc and Two-Step coating.  Filtered arc produces Pb layers \SIrange{2}{5}{\micro\metre} in thickness, but with the presence of spherical protrusions up to \SI{30}{\micro\metre} in diameter.  The Two-Step process generated thicker films up to 18 microns, with surface features up to \SI{50}{\micro\metre} in diameter reduced significantly following exposure to the Ar plasma in the second step.  XRD data implies a higher level of crystallinity from the Two-Step process, predominantly Pb(111), Pb(222) and Pb(333).

\section{\label{sec:WG5}WG5: Semiconductor Photocathodes}

Working Group 5 focused on the development and application of alkali--antimonide photocathodes and their use in accelerators.  It was convened by Maud Baylac.

\subsection{\label{WG5-1}Multi--alkali Antimonide Photocathodes for Highly Brilliant Electron Beams}

The first presentation of the semiconductor session was given by Chen WANG (HZB, Germany) on multi--alkali antimonide photocathodes for highly brilliant electron beams.  The Sealab/bERLinPro project aims to build a SRF gun with multi--alkali antimonides to reach high QE and long operation lifetime.  NaKSb photocathodes on a Mo substrate were produced in a UHV preparation chamber at HZB.  The influence of the deposition parameters was studied with XPS and QE measurements.  The correlation between chemical composition and QE values are discussed.  A hybrid deposition (combination of sequential and co--deposition) proved to produce the best performance.

\subsection{\label{WG5-2}Characterisation at Daresbury Laboratory of CsTe Photocathodes Grown at CERN}

Lee Jones (STFC Daresbury Laboratory, UK) talked on the performance characterization at Daresbury Laboratory of CsTe photocathodes manufactured at CERN. In the framework of a future UK X--FEL,  high--performance photocathodes are studied: surface characteristics are analyzed as they limit the achievable electron beam quality (emittance, brightness, energy spread). Detailed analysis of CsTe samples produced at CERN were performed with XPS, STM and the TESS spectrometer to evaluate the MTE. Multiple chemical species were found to be acting as photoemitters (Cs\textsubscript{x}Te\textsubscript{y}, Cs and CsO), and correlations between the measured surface characteristics and the MTE values for different wavelengths were shown. 

\subsection{\label{WG5-3}Cs\textsubscript{2}Te\textbackslash KCsSb in Gun Operation at PITZ}

The third talk, dedicated to development of multi--alkali antimonides photocathodes for high brightness photoinjectors, was given by Sandeep MOHANTY (DESY, Germany). For the KSbCs photocathodes synthesized at INFN--LASA, different factors affecting the growth procedure, such as substrate temperature and deposition rate, were detailed.  The photocathodes tested in the PITZ RF gun at DESY under high gradients showed promising results (QE, thermal emittance, and response time) despite high dark current and short operational lifetime.  Density--Functional Theory calculations were done with Quantum ESPRESSO.  NaKSb(Cs) exhibited several--year lifetime and resistance up to \SI{120}{\celsius} without degradation. 

\subsection{\label{WG5-4}Operation of Cs\textsubscript{2}Te in SRF gun\\for THz User Shifts}

The session concluded with Rong Xiang (HZDR, Germany) reporting on the operation of \cste{} in SRF gun for ELBE. For THz applications at the ELBE facility, the SRF gun--II operates with \cste{} photocathodes. After deposition on a Cu substrate, photocathodes were loaded and tested in the SRF gun. The evolution of QE during operation was analyzed and XPS measurement were done. It was shown that \cste{} photocathodes perform well in the SRF gun in terms of QE and charge lifetime (> \SI{10}{\coulomb}) with an acceptable level of dark current. It was underlined that a dedicated RF start up procedure is essential to avoid multipacting and thus preserve the cathode surface.

\section{\label{sec:ERLEurope}ERL Roadmap for Europe}

Boris Militsyn (STFC Daresbury Laboratory, UK) gave a review of the accelerator R\&D roadmap on Energy Recovery Linacs (ERL) and the significance of photocathode research in this field.  Within the 5 pillars of the European strategy for particle physics (\href{https://europeanstrategy.cern/}{https://europeanstrategy.cern/}), ERLs aim to push the energy frontier for collider applications: they are foreseen for FCC--eh and LHeC.  On the pathway to the demonstration of LHeC at the GW level, the PERLE project pursues multi--turn, multi--MW operation.  For these ERL applications, electron injectors must generate high average current (10s to 100s mA).  The presentation included a review of the state--of--the--art in the fields of DC, NCRF and SRF guns, extending to the major R\&D topics to be pursued such as green photocathodes, laser systems and novel gun technologies.

\section{\label{sec:WG6}WG6: Photocathode Theory}

Working Group 6 focused on the construction, development and testing of theoretical models supporting photoemission from photocathodes.  It was convened by Tim Noakes.

\subsection{\label{WG6-1}Monte Carlo Transverse Emittance and\\Quantum Efficiency Study on Cs\textsubscript{2}Te}

The first talk of the theory session was given by Gowri Adhikari (SLAC, USA) and concerned the theoretical modelling of Caesium Telluride photocathodes using Monte Carlo techniques.  The model focused on the ideal orthorhombic stoichiometric Cs\textsubscript{2}Te material, with the band structure and projected density of states determined for the bulk structure using density functional theory (DFT). This was then used in the three step photoemission model. The initial energy distribution was taken from the valence band distribution with the depth dependence calculated for the first \SI{15}{\nano\metre}. Particular attention was paid to the various scattering phenomena during electron transport to determine the distribution in the conduction band. The quantum efficiency and thermal emittance could then be calculated and both agreed well with experimental data if an assumed electron effective mass of 0.25 was used. Unusually for this type of calculation, a response time was also extracted from the simulations which indicated a value around \SI{200}{\femto\second}.

\subsection{\label{WG6-2}Monte Carlo Simulations of\\Electron Photoemission from\\Plasmon--enhanced Bialkali Photocathode}

The second talk of the session was given by Zenggong Jiang (Shanghai Advanced Research Institute, China). This talk concerned Monte Carlo simulations of photoemission from plasmon--enhanced bi--alkali photocathodes. The motivation of this study is to increase the adsorption of light in the surface region of the photocathode, both increasing QE and improving the response time. Surface plasmon adsorption can be enhanced by introducing a regular array of features into the surface with appropriate dimensions to match the wavelength. Three step photoemission was simulated. For a series of feature sizes, QE was enhanced by between 2 -- 3 times as a result of increased adsorption without any increase in the calculated emittance. Future work is planned to fabricate test structures to assess this enhancement technique experimentally.

\subsection{\label{WG6-3}Exploring Cesium–-Tellurium Phase Space\\via High--throughput Calculations beyond\\Semi--local Density--functional Theory}

The third talk was presented by Holger--Dietrich Sassnik (University of Oldenburg, Germany) where they have an active collaboration with experimentalists at HZB. The talk was about density functional theory calculations of binary Caesium based compounds (Cs\textsubscript{3}Sb and \cste{}) using high throughput methods. By using these methods the group were able to look at a whole range of different compositional and structural compounds in the two binary systems and establish the most stable ones. For the Cs-Te system they also carried out simulations taking into account the surface and evaluating the stability of different terminations ad determined their electronic properties.

\subsection{\label{WG6-4}Ultrafast Sub--threshold\\One-photon Photoemission}

The final talk of the session came from Andreas Schroeder (University of Illinois at Chicago, USA). This was a comparison of a new theoretical model for single photon photoemission with data collected using their QE and MTE measurement equipment. This experimental set up was described where the laser system has been upgraded to increase power in the UV region. In the model, the temperature of hot electrons and the ensuing Boltzmann tail in the energy distribution give rise to significant electrons that are able to access higher energy states within the conduction band that are in excess of the vacuum level. Hence there is no requirement for more than one photon to produce photoemission even below the work function. A comparison of data (QE and MTE) with theory for systems including Cu(111), GaSb(001), N--doped diamond(001) and Ag(111) demonstrated the general applicability of the approach to both metal and semiconductor photocathodes.

\section{\label{sec:WG7}WG7: Advanced Photocathode Characterization}

The last session of the EWPAA workshop in 2022 was dedicated to advanced photocathode characterization.  It was convened by Julius K\"uhn.

\subsection{\label{WG7-1}The ASU photocathode research lab}

Siddarth Karkare (Arizona State University, USA) talked about his Photoemission and Bright Beams Laboratory at ASU.  An ideal photocathode lab would consider photon and (cryogenic)--temperature dependence, surface preparation, photoemission characterization and tests in an electron gun. Everything has to be carried out in--situ under excellent extreme UHV conditions to avoid degradation of high QE Cs-based photocathodes. The system at ASU will be equipped with to growth chambers, one for alkali--antimonide thin film growth and one with RHEED and EDX capabilities to enable epitaxial grown films. Surface characterization can be done by UHV--AFM, LEED or Auger electron spectroscopy. For MTE-measurements and Electron Energy Analyzer with sub--meV resolution is available. A Photoemission Electron Microscope will allow to study nano--scale electron source, cathode emission uniformity and photonics/plasmonics integrated photocathodes. The system will also be equipped with an \SI{200}{\kV} DC electron gun with an acceleration gradient of \SI[per-mode=symbol]{8}{\mega\volt\per\metre} at \SI{200}{\kV} and a beam line for 4D space mapping and a \SI{3.0}{\giga\hertz} deflection cavity for temporal response measurements. The systems is supported by a tunable wavelength laser and cryocooling capabilities.

\subsection{\label{WG7-2}Photocathode Response Time Measurement}

The second talk of this session was given by Gregor Loisc (DESY Zeuthen, Germany) on behalf of the PITZ and INFN Milano cathode groups. He presented his results on direct response time measurements on semiconductor (and metal) photocathodes by measuring the bunch lengthening during the photoemission process. The measurement setup at the PITZ test facility is capable of studying the bunch length at low charge and short bunches with high TDS, to identify the RF--contribution to the bunch length via two--bunch probe and to disentangle cathode response time from the laser shape. The shortest response time for \cste{} measured was \SI{184 \pm 42}{\femto\second} and the longest 
\SI{257\pm 41}{\fs}. No aging effects and no dependence on the QE related of the cathode response time was observed. For an Au cathode a response time of \SI{183\pm 10}{\fs} was measured. The simulation results of the emission process show reasonable agreement.

\subsection{\label{WG7-3}Resolving Surface Chemical States of p-GaN:Cs Photocathodes by XPS Analysis}

The third talk of this session was given by Jana Schaber (Helmholtz Zentrum Dresden--Rossendorf, Germany) on resolving the surface chemical states of p-GaN:Cs photocathodes by in--situ photoelectron spectroscopy. A preparation chamber at low \SI{10e-10}{\milli\bar} vacuum level is equipped with a \SI{310}{\nm} UV--LED with a power of \SI{50}{\uW} and an XPS system. P-GaN was investigated as a potential photocathode material which can be used in several cycles whereby one cycle is defined as thermal cleaning, Cs--activation and QE--decay. It was found that Carbon an Oxygen are incorporated in the material by XPS and that Argon ion sputtering causes surface damage. The Cs activation step was also studied by XPS. QE values between \SIrange[range-units = single, range-phrase=--]{7}{12}{\percent} were achieved in the range of \SIrange[range-units = single, range-phrase=--]{400}{500}{\degreeCelsius}. p-GaN will be considered to be operated in the gun, but the cathode holder design still remains challenging.

\subsection{\label{WG7-4}High Crystalline Cesium Telluride Photocathode on Atomically--thin Graphene with Co--deposition}

The fourth talk was given by Mengjia Gaowei (Brookhaven National Laboratory, USA) on high crystalline cesium telluride photocathode on atomically--thin graphene via co--deposition.  At BNL a preparation system equipped with evaporators for alkali metals, Sb and Te, and with QCM, RHEED and QE-setup is operated which can be used at the NSLS--II light source doing XRD, XRR, XRF during the growth of the photocathode. Cesium telluride was grown via co--deposition and characterized by XRD online monitoring and XRR after growth. The growth recipe was optimized to achieve the composition of \cste{} on Si. The co--deposition incorporates more Cs than the sequential growth and leads to better crystallization, with low surface roughness and higher QE at \SI{266}{\nm}. Furthermore the growth of \cste{} on graphene/Si was studied, showing that the nucleation on graphene is more oriented than that on the Si substrate and the film is textured in the early stage of the growth. The use of graphene as a substrate for cesium telluride is a promising route to produce cathode film with better crystallinity and better cathode performance.

\subsection{\label{WG7-5}Molecular Beam Epitaxy of Cs--Sb Thin Films: Structure--oriented Growth of\\High Efficiency Photocathodes}

The fifth talk of the session was given by Alice Galdi (University of Salerno, Italy and CLASSE, Cornell University, USA) on the molecular beam epitaxy of Cs--Sb thin films to grow structure--oriented high efficiency photocathodes. Cs\textsubscript{3}Sb photocathodes have been grown epitaxially to control roughness and the surface potential by orientation control. As substrate for epitaxial growth 3C--SiC was identified. Reflection High Energy Electron Diffraction (RHEED) assisted MBE has been carried out to follow the crystallization process during thin film growth. RHEED can be operated in real time with sub--monolayer sensitivity and gives information about the surface roughness and crystallinity. The growth procedure has been optimized for single crystalline Cs\textsubscript{3}Sb.

\subsection{\label{WG7-6}Stoichiometry Control and Automated Growth\\of Alkali Antimonide Photocathode Films\\by Molecular Beam Deposition}

The last talk of the session was given by Vitaly Pavlenko (Los Alamos National Laboratory, USA) and focussed on stoichiometry control and automated growth of alkali antimonide photocathode films by molecular beam deposition. A method of film growth with support of a feedback loop that controls stoichiometry in real time is presented.  Photoemissive properties exhibit a distinct dependence on the stoichiometric composition that depends on the ratio of the incident fluxes. The results were obtained on Cs\textsubscript{3}Sb but are expected to be relevant to other alkali antimonides and tellurides.

\section{\label{sec:Conclusions}Conclusions}

This has been the first in-person Photocathode Workshop since the Covid-19 Pandemic. More than 40 participants attended the workshop, both from Europe and from the USA. 

We also greatly appreciated the willingness of colleagues from America and China who have endured the many hours of time zone difference in order to be present remotely and give their contribution to the success of the Workshop.

We thank all the Workshop participants for their contributions to very fruitful and stimulating discussions. 

The Local Organizer Committee thanks also the EWPAA Scientific Program Members for their support and help in organizing and achieving such a successful event.

The next workshop in the European Workshop On Photocathodes for Accelerator Applications series will be hosted by the Helmholtz Zentrum Dresden (HZDR) in the middle of 2024.




\end{document}